\documentclass[%
 reprint,
 amsmath,amssymb,
 aps,
]{revtex4-1}

\usepackage{graphicx}
\usepackage{dcolumn}
\usepackage{bm}

\begin{document}

\preprint{APS/123-QED}

\title{QUANTUM IDENTITIES FOR THE ACTION}

\author{E. Gozzi}
\affiliation{Department of Physics ,
Theoretical Section, University of Trieste, Strada Costiera 11, 34151 Trieste, Italy.\\
Istituto Nazionale di Fisica Nucleare, Sezione di Trieste, Italy.}





\date{\today}

\begin{abstract}
In this paper we derive various identities involving the {\it action} functional which enters the path-integral formulation of quantum mechanics. They provide  some kind of generalisations of the \break Ehrenfest theorem giving correlations between powers of the action  and its functional derivatives.
\begin{description}
\item[PACS numbers] 03.65.-W, 03.65.Db
\end{description}
\end{abstract}

\maketitle


\section{\label{sec:level1}Introduction}
The  {\it action} has been a central object of classical and quantum mechanics.  It was introduced by  Lagrange \cite{lagr}  in  classical mechanics and  he proved  that, by minimising  it,  the classical paths are recovered.  In quantum mechanics its role was discovered by R.P.Feynman \cite{fey} in his ph.d thesis in 1942. He proved that the action  is the  weight that each path must have in order to get the whole set of quantum effects  we know. The fact that  the action  had a role also in quantum mechanics was
clear already from the work of Max Planck whose energy quantisation involved a constant, the $\hbar$,
with the same  dimension of the action . 

\par 
We have the feeling anyhow that, especially at the quantum level, there are still things to discover regarding the action whose role is so central.

\par In order to make some progress concerning this issue,  in this paper we derive   some universal identities
involving the action . They are generalisations of the Ehrenfest theorem regarding  correlations between powers of the action  and its   higher derivatives. We had not found these kind of identities before in the literature so we thought it was worth to present them, hoping that someone can bring the project further.

\section{\label{sec:level1}The Action  In Classical and Quantum Mechanics}

Let us start by providing the definition of action . 
Given a path $x(t)$ in configuration space, the action  
$S[x(t)]$ is a {\it functional}  of the  trajectory $x(t)$ defined as:
\begin{equation}
S[x(t)]\equiv \int_{t_1}^{t_2} dt \Bigg ({1\over 2}m {\dot x(t)^{2}}-V(x(t))\Bigg ),
\label{due-uno}
 \end{equation} 
where the quantity inside the integral is the Lagrangian and the integral is evaluated between well defined limits.
The $S[x(t)]$ in eq.(\ref{due-uno})  is a {\it functional}  since once you give a {\it function} (the trajectory $x(t)$ ) the outcome $S[x(t)]$ is a number.
  
In classical mechanics Lagrange \cite{lagr} discovered that the classical trajectories  can be obtained
from the equation:
\begin{equation}
{\delta S[x(t)]\over \delta x(t)}=0,
\label{due-due}
 \end{equation} 
where the symbol ${\delta ()\over \delta x(t)}$ indicates the functional derivative.
The equation (\ref{due-due}) is a sort of minimisation of the action so some often say that {"\it the classical trajectories are those  which minimise the action"} . 

In quantum mechanics, instead,  the central objects are the transition amplitudes
$\langle {x_1,t_1} | {x_2,t_2}\rangle $. Feynman \cite{fey} proved that these amplitudes can be written as

\begin{equation}
\langle {x_1,t_1} | {x_2,t_2}\rangle=\int {\mathcal D} x(t) \, exp\,{i\over \hbar} S[x(t)]
\label{due-tre}
 \end{equation} 
 where  the $\int {\mathcal D} x(t) $ indicates the "{\it functional}" integration over all the trajectories between $(x_1,t_1)$ and $(x_2,t_2)$. The equation above is also called the {\it path-integral } version of quantum mechanics. Note that in (\ref{due-tre})  each trajectory $x(t)$ has a weight $ exp \,{i\over \hbar} S[x(t)]$ and this shows the role that the action has in quantum mechanics. If one would like to be more rigorous \cite{glimjaf}  the symbol  $\int {\mathcal D} x(t) $ should be replaced by the Wiener measure which contains the kinetic term and the action replaced by the integral of the potential. We will avoid this more rigorous approach in this paper in order to better bring to light the role of the action and  we shall stick to the original Feynman approach \cite{fey}.  We will call in the rest of the paper the symbol $\int {\mathcal D} x(t) $ as a measure even if it is not actually a proper  measure.
 
Expression (\ref{due-tre}) can be generalised to be the transition among general bra-state $\langle\phi |$ at time $t_1$ and ket-state $|\psi\rangle$ at time $t_2$ or between the vacuum of the theory etc. etc. In general if we sandwich a quantum  observable, whose  form is the one of a functional of $x(t)$ i.e ${\mathcal O}[x(t)]$,  between those states    and take the average, its path integral expression will be the following:
 
\begin{equation} 
\langle {\mathcal O}\rangle=\int {\mathcal D} x(t)  {\mathcal O}[x(t)]  \, exp \,{i\over \hbar} S[x(t)]
\label{due-quattro}
 \end{equation} 
 
 \section{\label{sec:level1}Quantum Identities for the Action}
 
  Since  in eq.(\ref{due-quattro})  we are integrating over paths, we can shift them, i.e. 
  \begin{equation} 
  {\mathcal D} x(t)={\mathcal D} [x(t)+\eta(t)]
 \label{tre-uno}
 \end{equation} 
 where $\eta(t)$ is an arbitrary small shift which respects the boundary conditions imposed on the paths $x(t)$. In the derivation which  follows we will reproduce  step by step the analog derivation contained in ref. \cite{schu}. The measure of integration  does not change under the shift and  $\langle {\mathcal O}\rangle$ does not change either because it does not depend on $x(t)$, hence  from eq.(\ref{due-quattro}), we get the following relation:
 \begin{equation}
 \begin{split}
 \int {\mathcal D} x(t)  \Bigl ({\mathcal O}[x(t) &+\eta(t)]  \, exp \,{i\over \hbar} S[x(t)+\eta(t)]+\\
 &-  {\mathcal O}[x(t)]  \, exp \,{i\over \hbar} S[x(t)]\Bigr)=0
 \label{tre-due}
 \end{split}
 \end{equation}
 
 Expanding in $\eta(t)$ the LHS of the equation above, we obtain:
 
 \begin{equation}
 \int {\mathcal D} x(t)\int d\sigma \Bigl [{\delta {\mathcal O}\over \delta x(\sigma)}+{i\over\hbar}{\mathcal O}{\delta S\over \delta x(\sigma)}\Bigr]\,\Bigl [exp{i\over\hbar}S\Bigr]\,\eta(\sigma)+O(\eta^2)
 \label{tre-tre}
 \end{equation}
 Let us remember that  $\eta(\sigma)$ is small but arbitrary  and so, combing eq.(\ref{tre-tre}) with 
 eq.(\ref{tre-due}), we get:
 
 \begin{equation}
 \left\langle{\delta{\mathcal O}\over \delta x(\sigma)}\right\rangle=-{i\over\hbar}\left\langle {\mathcal O}{\delta S\over
 \delta x(\sigma)}\right\rangle.
 \label{tre-quattro}
 \end{equation}
 
 If we choose $\mathcal O$ to be  the identity or a constant  $k$, i.e  ${\mathcal O}= k$, equation (\ref{tre-quattro}) gives:
 \begin{equation}
 \left\langle {\delta S\over
 \delta x(\sigma)}\right\rangle=0.
 \label{tre-cinque}
 \end{equation}
 This is nothing else than the well-know Ehrenfest theorem of quantum mechanics.
 
 Next let us choose the  ${\mathcal O}$ to be the action  $S[x(t)]$
. Note that in $S$ we have inserted an arbitrary trajectory $[x(t)]$ which anyhow is integrated in t inside the action so there is no explicit dependence on t in ${\mathcal O}$ . Inserting this into the general equation (\ref{tre-quattro}) we obtain:
  \begin{equation}
  \left\langle {\delta S\over
 \delta x(\sigma)}\right\rangle= -{i\over\hbar} \left\langle S[x(t)]{\delta S\over \delta x(\sigma)}\right\rangle.
 \label{tre-sei}
 \end{equation}
 The LHS of this equation is zero because of eq.(\ref{tre-cinque}), so the relation (\ref{tre-sei}) is turned into:
  \begin{equation}
  \left\langle S[x(t)]{\delta S\over \delta x(\sigma)}\right\rangle=0.
  \label{tre-sette}
 \end{equation}
 In the appendix of this paper we will give an explicit derivation of the relation above for both the free particle and the harmonic oscillator using the discretise form of the path-integral \cite{fey}.
 
 We can now continue by choosing $\mathcal {O}$ in the general equation (\ref{tre-quattro}) to be $S^2[x(t)]$, and we would get:
 \begin{equation}
 \left\langle{\delta{S^2}\over \delta x(\sigma)}\right\rangle=
 -{i\over\hbar}\left\langle {S^2[x(t)]}{\delta S\over
 \delta x(\sigma)}\right\rangle.
 \label{tre-otto}
 \end{equation}
 
 The LHS of this equation is 
 $$2\left\langle S[x(t)]{\delta S\over \delta x(\sigma)}\right\rangle$$
 which is zero because of eq.(\ref{tre-sette}), so the relation (\ref{tre-otto}) gives: 
 
 \begin{equation}
 \left\langle S^2[x(t)]{\delta S\over
 \delta x(\sigma)}\right\rangle=0.
 \label{tre-nove}
 \end{equation}
 
 Next if we choose $\mathcal {O}$  to be $S^3[x(t)]$ we will get via the same procedure the relation:\begin{equation}
 \left\langle S^3[x(t)]{\delta S\over
 \delta x(\sigma)}\right\rangle=0.
 \label{tre-dieci}
 \end{equation}
 
 We can continue in this way and get in general the relation:
 
 \begin{equation}
 \left\langle S^n[x(t)]{\delta S\over
 \delta x(\sigma)}\right\rangle=0,
 \label{tre-undici}
 \end{equation}
 where $n$ is an arbitrary  integer value. The relations above depend explicitly on the instant of  time $\sigma$ but not on the  instant of time $t$ which is integrated inside the action . So the above relations are not properly correlations at different instants of time.
 
 \par
 Note that in the relations above the "correlation" between the equation  of motion ${\delta S[x(\sigma)]\over\delta x(\sigma)}$ and  the rest is zero only if that "something else" is the action  or its powers otherwise we would get eq.(\ref{tre-quattro}). So the relations (\ref{tre-undici}) are unique for the action .
 
 As the eq.(\ref{tre-undici}) above is valid  for any $n$ an immediate consequence is that also the following relation holds:
 \begin{equation}
 \left\langle \biggl[exp {i\over\hbar}S[x(t)]\biggr]{\delta S\over
 \delta x(\sigma)}\right\rangle=0.
 \label{tre-dodici}
 \end{equation}
 This is so because if we expand the $exp {i\over\hbar}S$ in powers of ${i\over\hbar}$ we get a set of terms identical to eq.(\ref{tre-undici}) each term having a different $n$.  The eq.(\ref{tre-dodici}) can be written in full path-integral form and we would get :
 
 \begin{equation}
 \int{\mathcal D}x(t)\,exp \, \, 2{i\over\hbar}S[x(t)]\,{\delta S\over
 \delta x(\sigma)}=0.
 \label{tre-dodici-a}
 \end{equation}
 
 The factor "2" , featuring  in the exponent, emerges  after combining the exponential of the path-integral weight with the exponential appearing in eq.(\ref{tre-dodici}). Actually from eq.(\ref{tre-undici}) we get that, by introducing an arbitrary parameter $\lambda$, the following relations would also hold: 
 
 \begin{equation}
{({i\lambda\over \hbar})^{n}\over n!}
 \left\langle S^n[x(t)]{\delta S\over
 \delta x(\sigma)}\right\rangle=0,
 \label{tre-undici-a}
 \end{equation}
 
 If we  sum up all these terms and write the full path-integral ,  we would get the following generalisation
 of  eq.(\ref{tre-dodici-a}): 
 
 \begin{equation}
 \int{\mathcal D}x(t)\,\Bigl\{exp \, {i(\lambda+1)\over\hbar}S[x(t)]\Bigr\}\,{\delta S\over
 \delta x(\sigma)}=0.
 \label{tre-dodici-b}
 \end{equation}
 
 This indicates that  we can somehow rescale  $\hbar$ and nothing changes. It is as if this was "classical physics" where nothing changes with $\hbar$. It is easy to give a different proof for the eq. (\ref
 {tre-dodici-b}) as follows. Let us define a new action :
 
 \begin{equation}
 {\widetilde S}=(\lambda +1) S
 \end{equation}
 
 then equation (\ref{tre-dodici-b}) can be rewritten as :
 
 \begin{equation}
 {1\over (\lambda+1)}
 \int{\mathcal D}x(t)\,exp \, {i\over\hbar}{\widetilde S}[x(t)]\,{\delta {\widetilde S}\over
 \delta x(\sigma)}=0.
 \label{tre-dodici-bb}
 \end{equation}
This relation holds being the analog of eq.(\ref{tre-cinque}).
 
 Let us now move on with other choices of the $\mathcal {O}$ in eq.(\ref{tre-quattro}). If we choose:

\begin{equation}
 \mathcal {O}= S[x(t)]^{-1}
 \label{tre-dodici-l}
 \end{equation} 
 
 we get , from (\ref{tre-quattro}) , the following  identity:
 
 \begin{equation}
 \left\langle S^{-2}{\delta S[x(t)]\over \delta x(\sigma)}\right\rangle={i\over\hbar}\left\langle S^{-1}{\delta S[x(t)]\over \delta x(\sigma)}\right\rangle.
 \label{tre-dodici-m}
 \end{equation}

If we now choose: 
\begin{equation}
 \mathcal {O}= S[x(t)]^{-2}
 \label{tre-dodici-n}
 \end{equation} 
 we obtain :
 
 \begin{equation}
 \left\langle 2S^{-3}{\delta S[x(t)]\over \delta x(\sigma)}\right\rangle={i\over\hbar}\left\langle S^{-2}{\delta S[x(t)]\over \delta x(\sigma)}\right\rangle.
 \label{tre-dodici-o}
 \end{equation}
 
 In general with the choice :
  \begin{equation}
 \mathcal {O}= S[x(t)]^{-n}
 \label{tre-dodici-r}
 \end{equation} 
  we get the relation:
 
 \begin{equation}
 \left\langle nS^{-n-1}{\delta S[x(t)]\over \delta x(\sigma)}\right\rangle={i\over\hbar}\left\langle S^{-n}{\delta S[x(t)]\over
 \delta x(\sigma)}\right\rangle.
 \label{tre-dodici-s}
 \end{equation}
 
 Notice that in all these relations the RHS and the LHS are very similar  with just the  power of $S$ differing by one on the two sides, while the rest identical. Somehow $h$ on the RHS replaces one of the $S$ featuring on the LHS.
 
\par
 Let us now move on with further choices for the $\mathcal {O}$ . In particular let us make the following choice:
 
 \begin{equation}
 \mathcal {O}={\delta S[x(t)]\over \delta x(\tau)}
 \label{tre-tredici}
 \end{equation}
 
 Note that differently than before now $\mathcal {O}$ depends on the time  parameter $\tau$
 
 Inserting this expression in  eq.(\ref{tre-quattro}) we get:
 \begin{equation}
 \left\langle{\delta^2 S[x(t)]\over \delta x(\tau)\delta x(\sigma)}\right\rangle=-{i\over\hbar}\left\langle {\delta S\over
 \delta x(\tau)}{\delta S\over
 \delta x(\sigma)}\right\rangle.
 \label{tre-tredici}
 \end{equation}
 
Note that the LHS of this relation contains the second variation of $S$, which is the crucial ingredient one
needs when considering the fluctuations around classical paths. It also plays a key role in many other topics  \cite{schu} like  the WKB approximation, the Lyaupunov exponents, etc. This relation links  this quantity to the correlation function among the equations of motion  entering  the RHS of eq.(\ref{tre-tredici})
\par
If we now continue with this procedure  and choose 

\begin{equation}
 \mathcal {O}={\delta^2 S[x(t)]\over \delta x(\tau)^2}
 \label{tre-quattordici}
 \end{equation}
 
 we would get from the equation (\ref{tre-quattro})
 
  \begin{equation}
 \left\langle{\delta^3 S[x(t)]\over \delta x(\sigma)\delta x(\tau)^2}\right\rangle=-{i\over\hbar}\left\langle {\delta^2 S\over
 \delta x(\tau)^2}{\delta S\over
 \delta x(\sigma)}\right\rangle.
 \label{tre-quindici}
 \end{equation}
 
 In general, making the following choice 
\begin{equation}
\mathcal {O}={\delta^{m-1}S[x(t)]\over \delta x(\tau)^{m-1}}
\label{tre-sedici}
\end{equation}
 
 we obtain from (\ref{tre-quattro}) the relation:
 \begin{equation}
\left\langle{\delta^{m} S[x(t)]\over \delta x(\sigma)\delta x(\tau)^{m-1}}\right\rangle=-{i\over\hbar}\left\langle {\delta^{m-1} S\over
\delta x(\tau)^{m-1}}{\delta S\over
\delta x(\sigma)}\right\rangle.
\label{tre-diciassette}
\end{equation}
\par

This indicates that expectation value of higher order derivatives of the action can be obtained from the correlations between lower order ones. 

 \section{\label{sec:level1}Conclusions and Outlook}
 
 The correlations among higher derivatives of the action,  that we have derived in the previous sections,  will be useful especially in field theory because they are not based on symmetries of a particular  model but are quite general and based only on the general structure of the path-integral. In field theory people always look for relations among correlation functions  and here we have several of them.

 Besides these applications  our plan for the future is to understand better the {\it physical} meaning of the action even at the classical level. We feel in fact that while for quantities like energy, angular momentum, etc their physical meaning is very clear, it is not so for the action. The reason is due to the "strange" {\it minus sign} between the kinetic and the potential terms in the Lagrangian. That minus sign is crucial in order to get the correct equations of motion from the minimisation of the action but, at the same time, it is the object that obscures the physical meaning of the Lagrangian. In optics the analog minimisation principle of mechanics is called Fermat principle \cite{toronto}. There the role of the action is taken by the "optical path length" which is the product of the distance among the two points in space between which the light travels and the index of refraction of the material in which the beam propagates. For sure this quantity has a clear physical meaning. In order to get an analog physical interpretation in classical mechanics, the Lagrangian should be interpreted (modulo a dimensional constant) as the product of the velocity times the index of refraction of a "would be" media
 where the particle travels. We do not like this. The direction to go must be different. The one we have been exploring   seems to tell us that the action (modulo a constant) is related to the   {\it entropy} of the Nelson stochastic process \cite{nelso}. This is a process that has been widely used in axiomatic field theory (see \cite{wight} and references therein). We hope to give more details in a future paper.

 \begin{acknowledgments}
The people I would like to thank for helpful   discussions would be too many to mention all of them and besides they know their names. This work has been supported  by INFN (IS: geosymqft , Naples and gruppo IV , Trieste ). \end{acknowledgments}

\bibliography{apssamp}

\begin{thebibliography}{20}

\bibitem{lagr}
J.L.Lagrange, "{\it Mecanique Analytique}", chez Veuve Desaint Libraire, Paris 1788.
\bibitem{fey} 
R,P.Feynman, "{\it Feynman's Thesis: a new approach to quantum mechanics}" World-Scientific, Singapore (2005); Rev.Mod.Phys. (20) 367 (1948).
\bibitem{glimjaf}
J.Glimm, A.Jaffe, "{\it Quantum Physics: A Functional Integral Point of View}",
Springer-Verlag, Berlin-Heidelberg-New York, (1981).
\bibitem{schu} L.S.Schulman, "{\it Techniques and Applications of Path Integration}" J.Wiley and Sons, New York (1981).
\bibitem{hib} R.P.Feynman and A.R.Hibbs, "{\it Quantum Mechanics and Path Integrals}", McGraw-Hill, New York (1965) .
\bibitem{enn} A.A.Abrikosov jr, E.Gozzi, D.Mauro, Ann.Phys.317 (2005) 24; E.Gozzi, E.Cattaruzza, C.Pagani, "{\it Path Integrala for Pedestrians}", World Scient. Publ., Singapore (2016).
\bibitem{RPF} R.P.Feynman. "{\it The Feynman Lectures on Physics}" Vol.2, Part 1. Addison-Wesley Publishing company, London, 1969.
\bibitem{toronto} M.Born and E.Wolf "{\it Principles of Optics"},  Cambridge University Press, 1999; C.Lanczos "{\it The Variational Principles of Mechanics"}, Toronto University Press, 1949.
\bibitem{nelso}
E.Nelson, Phys.Rev.150 (1966) 1079.
\bibitem{wight}
F.Guerra and P.Ruggiero, Phys.Rev.Lett. 31, 1022, (1973)
 \end{thebibliography}
  
\end{document}